# Exact WKB-like Formulae for the Energies by means of the Quantum Hamilton-Jacobi Equation

Mario Fusco Girard

Department of Physics "E. R. Caianiello"

University of Salerno, Italy

Electronic address: mfuscogirard@gmail.com

Abstract

It is shown that by means of the approach based on the Quantum Hamilton-Jacobi equation, it is possible to modify the WKB expressions for the energy levels of quantum systems, when incorrect, obtaining exact WKB-like formulae. This extends the results found in previous papers, where it was demonstrated that the QHJ method provides exact WKB-like expressions for the wave functions.

PACS numbers. 03.65, Ca

## Introduction

The well-known WKB approximation method [1] provides simple expressions for the quantum energy levels. In some cases, these formulae give the exact quantum values, in other cases the WKB values differ from these, and the differences can or not depend on the quantum number *n*, which is connected to the number of zeroes of the wave function.

The aim of this paper is to show that it is possible to explain why this happens, by means of the quantum Hamilton-Jacobi approach to Quantum Mechanics (QM) [2-7]. This is fully in the framework of the Copenhagen interpretation of QM, is equivalent to this latter, and allows to fully investigate the links between QM and the classical mechanics. This method permits also to modify the WKB formulae, when incorrect, in order to obtain similar but exact expressions for the energies.

The paper is organized as follows: in the Section 1, a brief review of the WKB derivation of the formulae for the energy levels is presented, together with a schematic introduction to the quantum Hamilton-Jacobi method. The reason why the WKB formulae are sometimes incorrect is explained. Afterwards, it is shown how the QHJ approach can be exploited to obtain the exact WKB-like formulae. In the Section 2, the method is applied to various quantum systems.

## **Section 1.**

1 a. *The WKB energies.*

The WKB method is exposed in almost all the books on QM [8]. Detailed modern reviews of the method (also named as JWKB or phase-integral method) are presented for instance in the Refs. [9-11]. Very briefly, the WKB formulae for the energies are obtained as follows. Let us consider a particle of mass m and energy *E*, moving along the x-axis under a potential *V(x)*, with two turning points $x_1 < x_2$, where *V(x$_i$)* = *E*.

Indicating as *p$_C$ (x, E)* the classical momentum

$$p_C(x, E) = \pm\sqrt{2m(E - V(x))} \qquad (1),$$

the study of the WKB wave function demonstrates that the semi classical energy *E* has to satisfy the condition

$$\int_{x_1}^{x_2} p_C(x, E)\, dx = \left(n + \frac{1}{2}\right)\pi\hbar \qquad n = 0, 1, 2... \qquad (2).$$

The left side of (2) can be computed as function of the energy, by integrating eq. (1) between the turning points:

$$\int_{x_1}^{x_2} p_C(x, E)\, dx = \int_{x_1}^{x_2} \sqrt{2m(E - V(x))}\, dx \qquad (3).$$

From (2) and (3) one obtains:

$$\int_{x_1}^{x_2} \sqrt{2m(E - V(x))}\, dx = \left(n + \frac{1}{2}\right)\pi\hbar \qquad (4).$$

By computing the integral (remembering that the turning points are functions of *E*) and extracting the energy, one gets the WKB expression for the semi classical energy levels, as a function of the quantum number *n* and of the parameters of the potential. The simplest example is the harmonic oscillator, for which the energy levels computed according to the WKB approach are:

$$E = \hbar\omega(n + \tfrac{1}{2}) \qquad (5),$$

and coincide with the exact quantum levels.

*1 b. Calculation of the energies by means of the Quantum Hamilton-Jacobi approach.*

The Quantum Hamilton-Jacobi Equation (QHJE) appears when the particle's wave function is searched for in the form:

$$\psi(x,E,t) = A e^{\tfrac{i}{\hbar}S(x,E,t)} \qquad (6),$$

where *S(x, E, t)* is a complex quantity, and *A* is a constant.

When eq. (6) is inserted into the Schrödinger equation:

$$i\hbar \frac{\partial \psi}{\partial t} = -\frac{\hbar^2}{2m}\frac{\partial^2 \psi}{\partial x^2} + V(x)\psi \qquad (7),$$

the QHJE results:

$$-\frac{\partial S}{\partial t} = \frac{1}{2m}\left(\left(\frac{\partial S}{\partial x}\right)^2 - i\hbar \frac{\partial^2 S}{\partial x^2}\right) - V(x) \qquad (8).$$

The time dependence can be separated, by writing:

$$S(x, E, t) = W(x, E) - E\,t \qquad (9),$$

Then, eq. (8) becomes the time-independent QHJE:

$$\frac{1}{2m}\left(\frac{dW}{dx}\right)^2 - \frac{i\hbar}{2m}\frac{d^2W}{dx^2} = E - V(x) \qquad (10).$$

Like *S(x, E, t)*, the function *W(x, E)* is in general a complex quantity.

In the following, the derivatives with respect to x will be indicated by means of primes (i.e. $\frac{dW}{dx} = W'(x,E)$). By setting $\hbar = 0$, the equations (8) and (10) become the classical time-dependent and time-independent Hamilton-Jacobi equations, respectively, whose solutions $S_C(x, E, t)$ and $W_C(x, E)$ are the Hamilton's principal and characteristic functions [12], also named action and abbreviated action, respectively [13].

Therefore, a solution *S(x, E, t)* of the QHJE (8) will be in analogy called the quantum Hamilton's principal function (or simply quantum action), and a solution *W(x, E)* of the eq. (10) will be called the quantum Hamilton's characteristic function (or quantum abbreviated action).

While we defer to the quoted references [2-7] for a detailed exposition of the QHJ approach, let us see how it is possible to apply it to find exact WKB-like formulae for the energy.

In the classical region between the turning points, we search for a solution of eq. (10) in the form:

$$W(x,E) = X(x,E) + iY(x,E) \qquad (11).$$

By inserting this into eq. (10) and separating the real and the imaginary parts, one obtains (in the following, the dependence of the various quantities on the energy $E$ will be often understood):

$$X'^2(x) - Y'^2(x) + \hbar Y''(x) = 2m(E - V(x)) \qquad (12),$$

$$X'(x)Y'(x) + \frac{1}{2}\hbar Y''(x) = 0 \qquad (13).$$

Last equation gives:

$$Y(x) = \hbar \text{Log}(\sqrt{X'(x)}) \qquad (14).$$

By putting (14) into (12), the following equation results:

$$X'^2(x) - \frac{3\hbar^2 X''^2(x)}{4X'^2(x)} + \frac{\hbar^2 X'''(x)}{2X'(x)} = 2m(E - V(x)) \qquad (15)$$

The right hand side of (15) is simply the square of the classical momentum $p_C(x)$.

This third order differential equation is rigorously equivalent to the Schrödinger equation [14, pag. 232].

$X'(x)$ and $Y'(x)$ are respectively the real and the imaginary part of the so-called quantum momentum function $p(x)$ [2, 3]:

$$p(x) = W'(x) = X'(x) + iY'(x) \qquad (16).$$

It is a complex function of the coordinate and the energy, not to be confused with the quantum operator momentum, which does not enter in this approach.

As demonstrated in [5], in the limit when $\hbar \to 0$ the imaginary part $Y(x)$ vanishes, the real part of the quantum abbreviated action $X(x)$ becomes the classical corresponding quantity $W_C(x)$, and the real part of the quantum momentum $X'(x)$ becomes the classical momentum $p_C(x)$. This is confirmed by the fact that, when $\hbar \to 0$, the eq. (15) reduces to

$$X'^2(x) = 2m(E - V(x)) \qquad (17),$$

i.e. the equation which links the square of the classical momentum to the energy.

In the QHJ approach, the wave functions between the turning points have the exact WKB-like expression:

$$\psi(x) = \frac{A}{\sqrt{X'(x)}} \text{Sin}\left(\frac{X(x)}{\hbar} + \frac{\pi}{4}\right) \qquad (18),$$

where $A$ is a constant. The semi classical WKB wave functions have the same form, but with the classical abbreviated action $W_C(x)$ instead of the real part of the quantum one $X(x)$, and the classical momentum $p_C(x)$ in the place of the real part of the quantum momentum function $X'(x)$.

The same considerations that in the WKB approach bring to eq. (2), when applied to the exact expression (18) lead to the condition:

$$\int_{x_1}^{x_2} X'(x)\,dx = (n + \tfrac{1}{2})\pi\hbar \qquad (19).$$

The l.h.s. of (15) can be written as:

$$X'^2(x)(1 + F(x)) \qquad (20),$$

where

$$F(x) = -\frac{3\hbar^2 X''^2(x)}{4X'^4(x)} + \frac{\hbar^2 X'''(x)}{2X'^3(x)} \qquad (21).$$

For all the potential V(x) so far investigated, the function F(x) has at least one zero $x_0$ between the turning points. There, the real part of the quantum momentum X'($x_0$) equals the classical momentum $p_C(x_0)$.

The square root of (15) is:

$$\sqrt{X'^2(x)(1 + F(x))} = \sqrt{2m(E - V(x))} \qquad (22).$$

Taking out X'(x) in the l.h.s. and expanding the square root near one of the zeroes of F(x), the last equation becomes:

$$X'(x)(1 + G(x)) = \sqrt{2m(E - V(x))} \qquad (23),$$

where $G(x)$ is the sum of the (converging) series:

$$G(x) = \tfrac{1}{2}F'(x_0)(x - x_0) + \tfrac{1}{8}(-F'^2(x_0) + 2F''(x_0))(x - x_0)^2 + \ldots \quad (24).$$

By integrating (23) between the turning points, one has:

$$\int_{x_1}^{x_2} X'(x)(1 + G(x))\,dx = \int_{x_1}^{x_2} X'(x)\,dx + R = \int_{x_1}^{x_2} \sqrt{2m(E - V(x))}\,dx \qquad (25),$$

where

$$R = \int_{x_1}^{x_2} X'(x) G(x)\,dx \qquad (26).$$

Finally, by inserting the condition (19), the last equality in (25) takes the form:

$$\int_{x_1}^{x_2} \sqrt{2m(E - V(x))}\,dx = (n + \tfrac{1}{2})\pi\hbar + R \qquad (27).$$

This is the exact version of the WKB formula (4), and is the main result of this paper. It implies that when the residual quantity R in (26) is different from zero, the WKB formula cannot give the exact quantum energy levels.

As shown in [5], the equation (15) can be analytically integrated for some systems (harmonic oscillator, hydrogen atom). Other systems require numerical computations [4, 6, 7]. The procedure requires to choose an eigenvalue E of the energy for the given potential V(x). Then the differential equation (15) is integrated as described in the quoted references, and the various quantities in the previous equations can so be computed. For all the potentials so far investigated, the functions X(x),

*F(x)* and *G(x)* are regular functions in the interval between the turning points. The residual quantity *R* in (26) can so be computed without problems. The computation is then repeated for different values of the energy *E*. Depending on the potential, three cases can occur:

1) *R* is equal to zero for each value of *n*;
2) *R* is different from zero and independent on *n* but dependent on some parameters of the potential (for instance the value of the angular momentum);
3) *R* is different from zero and dependent on *n* and some parameters.

In the first case, the formula WKB for the energies gives the exact quantum values.

In the second case, the WKB formula can be corrected to give the exact values, by taking into account the residual quantity *R*. The deduction of the corrected formula WKB-like requires only to add *R* in the r.h.s. of the eq. (4).

The same can be done in the third case, but now the correction is dependent on the quantum number *n* and other parameters.

The following section presents some examples for each case.

### Section 2

*2 a. The Harmonic and Morse oscillators*

When *R* is independent on *n* and equal to zero, the WKB formula for the energy is exact. In this case, the variation of the real part *X(x)* of the quantum abbreviated action between the turning points:

$$X(x_2) - X(x_1) = \int_{x_1}^{x_2} X'(x)\, dx \qquad (28)$$

equals the variation of the corresponding classical quantity, the classical Hamilton's characteristic function at the same energy:

$$W_C(x_2) - W_C(x_1) = \int_{x_1}^{x_2} p_C(x)\, dx \qquad (29).$$

The simplest example of such systems is the harmonic oscillator. This case can be treated both analytically and numerically. In Fig. 1, some results for it are reported, calculated by putting equal to 1 the mass m and the Planck's constant $\hbar$ (as in the rest of the paper) and with the angular frequency ω = 1. Moreover, all the computations presented were done by setting *X(x₁)* = 0. In the left box are plotted the real part *X(x)* of the quantum action for the *n* = 2 state (blue, continuous line), together with the corresponding classical function *W_C (x)* (orange, dashed). The quantum function follows waving the curve of the classical quantity. The two curves start and end in the same points, which shows that the residual quantity $R = 0$, so that the WKB energy levels are exact.

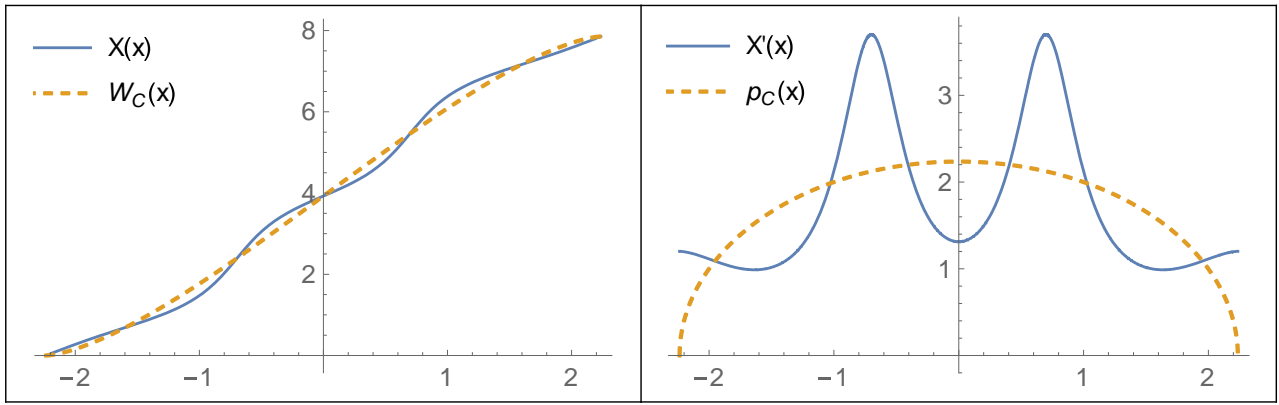

Fig.1.

In the right box are plotted, for the same state, the real part $X'(x)$ of the quantum momentum, together with the corresponding classical momentum $p_C(x)$. The intersections of these two lines correspond to the zeroes of the function $F(x)$ in eq. (21). The function $X'(x)$ presents peaks of finite heights in correspondence of the zeroes of the wave function. The peaks' height is proportional to $\sqrt{n}$ and while increasing the value of $n$, the peaks become higher and sharper. This behavior of $X'(x)$ is common to all the other potentials, and the same holds for the oscillations of $X(x)$ around its classical counterpart $W_C(x)$.

In Fig. 2 are plotted, for the same state $n = 2$ of the harmonic oscillator, the function $F(x)$ in the eq. (21) (left box), and the function $G(x)$ in the eq. (24).

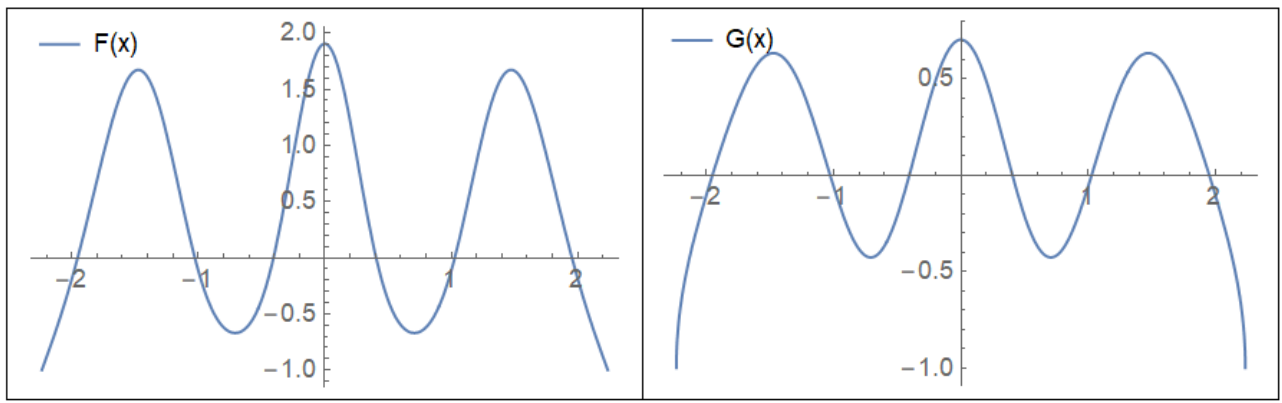

Fig.2

The Fig. 3 reports the result of an analogous computation for the state $n = 2$ of the Morse system, with the potential

$$V(x) = V_0(\text{Exp}(-2ax) - 2\text{Exp}(-ax)) \qquad (30).$$

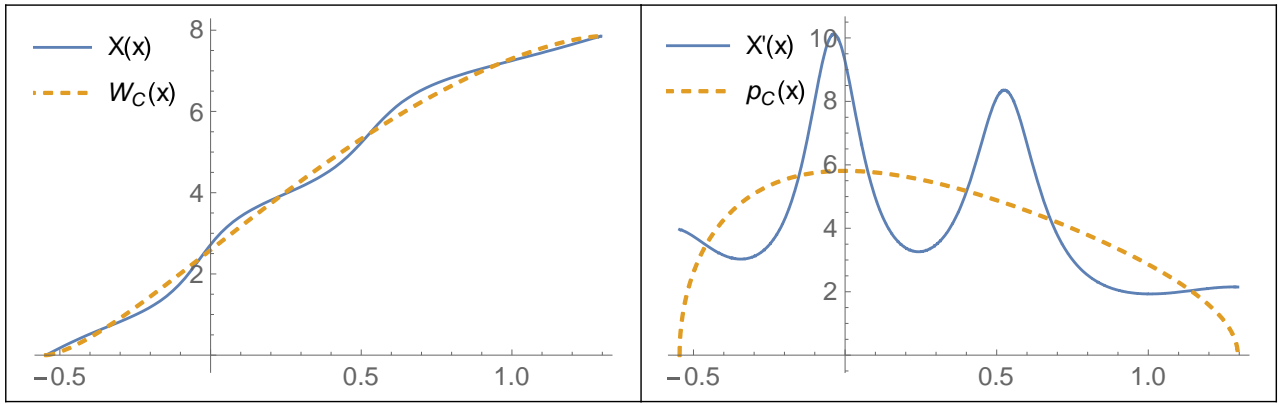

Fig.3

The computations were done with the parameters $a = 1$ and $V_0 = 32$. For this system too, the residue $R$ is zero. Also in this case, the WKB formula for the energies

$$E_{\text{WKB}} = -V_0 \left(1 - \frac{a\hbar}{\sqrt{2mV_0}}\left(n+\frac{1}{2}\right)\right)^2 \qquad (31)$$

coincides with the quantum one.

*2b) R independent on n but depending on some parameter.*

The most important example is the hydrogen atom, with the potential (coulomb + centrifugal):

$$V(r) = -\frac{e^2}{r} + \frac{l(l+1)}{2mr^2} \qquad (32).$$

The residual quantity $R$ for this system depends on the value of the azimuthal quantum number $l$.

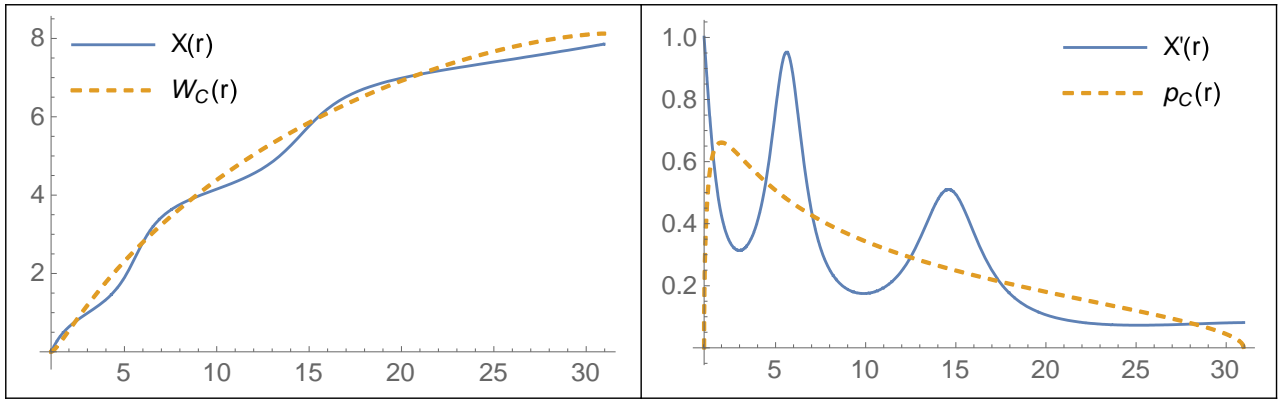

Fig. 4

The Fig. 4 reports some results for this system, as functions of the radius r. The figure refers to the values $n = 4$, $l = 1$, with the electric charge $e = 1$. From the left box of the figure, it is clear that the variation of the quantum abbreviated action differs from that of the classical action (the two curves start from the same point, but end differently), so that the residual quantity $R$ is different from zero. Moreover, the computations show that it does not depend on $n$, but only on the azimuthal number $l$. Some values for $R$ are given in the table 1.

| n | l | R |
|---|---|---|
| 1 | 0 | π/2 |
| 2 | 1 | 0.269506 |
| 3 | 1 | 0.269506 |
| 3 | 2 | 0.158683 |
| 4 | 1 | 0.269506 |
| 4 | 2 | 0.158683 |
| 4 | 3 | 0.112778 |
| 5 | 4 | 0.0875375 |
| 5 | 3 | 0.112778 |
| 5 | 2 | 0.158683 |
| 5 | 1 | 0.269506 |
| 6 | 5 | 0.071548 |
| 6 | 4 | 0.0875375 |
| 6 | 3 | 0.112778 |
| 6 | 2 | 0.158683 |
| 6 | 1 | 0.269506 |

Table 1.

The WKB formula for the energies is [15]:

$$E_{\text{WKB}} = -\frac{me^4}{2\hbar^2} \frac{1}{(n_r+\frac{1}{2}+\sqrt{l(l+1)})^2} \qquad (33),$$

where $n_r$ is the radial quantum number.

The corrected WKB-like formula is:

$$E_{\text{WKB,corrected}} = -\frac{me^4}{2\hbar^2} \frac{1}{(n_r+\frac{1}{2}+\frac{R}{\pi}+\sqrt{l(l+1)})^2} \quad (34).$$

With the values of *R* computed as described, the corrected formula gives the same energy levels as the quantum formula, where *n* is the principal quantum number:

$$E = -\frac{me^4}{2\hbar^2}\frac{1}{n^2} \quad (35).$$

The radial quantum number $n_r$ in (34) can be expressed in terms of *n* and *l*:

$$n_r = n - l - 1 \quad (36).$$

Then, the eq. (34) becomes (35) if R satisfies the relation:

$$R = \pi(l - \sqrt{l(l+1.)} + 1/2) \quad (37).$$

It is easy to check that (37) reproduces the values of *R* computed by means of the QHJE (15), reported in the Table 1.

### 2 c) The potential $ctg^2$

Another system of this type has the potential:

$$V(x) = V_0 \, \text{ctg}^2(\pi x/a) \quad (38).$$

In Fig. 5 are reported the usual quantities for this system. The curves are computed with $V_0$ = 1 and *a* = π. The figure refers to the *n* = 3 state, with the energy *E* = 7.

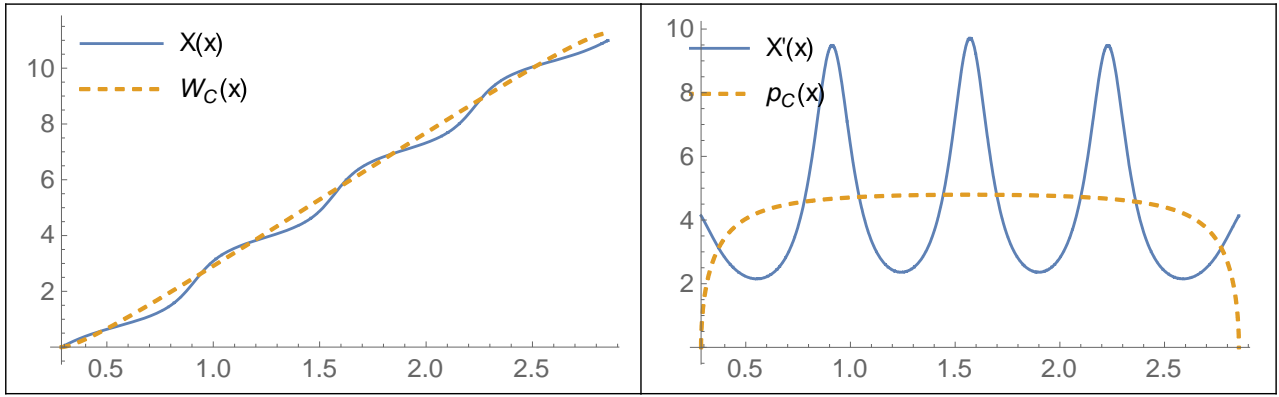

Fig. 5

The residual quantity *R* for this system is independent on *n*, and depends on the parameters $V_0$ and *a*. With the indicated values of the parameters, the residual quantity is *R*= 0.269506 and does not depend on *n*.

The WKB formula for the energies is [16]:

$$E_{\text{WKB}} = \frac{\pi^2 \hbar^2}{2ma^2} \left( \sqrt{\frac{2 \cdot ma^2 V_0}{\pi^2 \hbar^2}} + \left(n + \frac{1}{2}\right) \right)^2 - V_0 \qquad n = 0, 1, 2\ldots \qquad (39).$$

Therefore, the corrected formula is:

$$E_{\text{WKB,corrected}} = \frac{\pi^2 \hbar^2}{2ma^2} \left( \sqrt{\frac{2\, ma^2 V_0}{\pi^2 \hbar^2}} + \left(n + \frac{1}{2} + \frac{R}{\pi}\right) \right)^2 - V_0 \qquad n = 0, 1, 2\ldots \qquad (40).$$

The exact quantum formula for the energy is [16]:

$$E = (n^2 + 4\,n\,\lambda - 2\,\lambda)\frac{\pi^2 \hbar^2}{2ma^2} \qquad n = 1,2,3\ldots \qquad (41),$$

where

$$\lambda = \frac{1}{4}\left(\sqrt{\frac{8mV_0\, a^2}{\pi^2 \hbar^2} + 1} - 1\right) \qquad (42).$$

For

For each value of $n$, the eqs. (40) and (41) give the same energies. From these equations, is possible to calculate $R$ as a function of the parameters $V_0$ and $a$, but the resulting expression is a bit cumbersome and will not be reported here.

### 2.d  The purely quartic oscillator

An example of a potential of the third type is the purely quartic oscillator. The potential is:

$$V(x) = -\frac{x^4}{a} \qquad (43).$$

The Fig. 6 refers to this system, with $a = 1$. The state has $n = 4$, with the energy $E = 10.244308$.

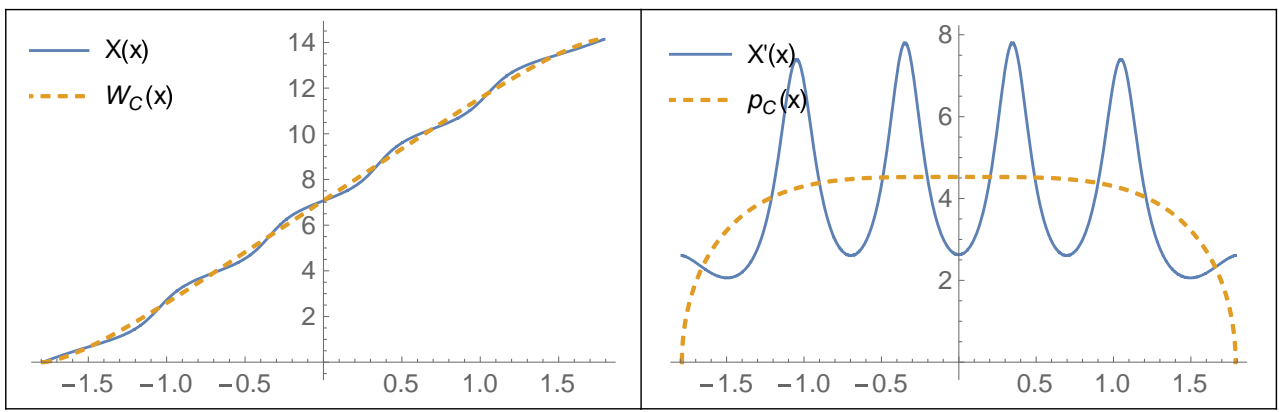

Fig. 6

The WKB formula for the energies is:

$$E_{\text{WKB}} = \left(\frac{\sqrt{\pi}\hbar 4 a^{\frac{1}{4}} \Gamma[\frac{7}{4}]}{\sqrt{2m}\,\Gamma[\frac{1}{4}]}\right)^{\frac{4}{3}} \left(n+\frac{1}{2}\right)^{\frac{4}{3}} \tag{44}$$

The residual quantity R depends on n, and its first few values, computed for a = 1, are reported in the Table 2.

| n | E | R |
|---|---|---|
| 0 | 0.667986 | 0.255796 |
| 1 | 2.393644 | 0.044912 |
| 2 | 4.696795 | 0.0331155 |
| 3 | 7.335730 | 0.0235851 |
| 4 | 10.244308 | 0.0184221 |

Table 2.

The corrected WKB formula is:

$$E_{\text{WKB,corrected}} = \left(\frac{\sqrt{\pi}\hbar 4 a^{\frac{1}{4}} \Gamma[\frac{7}{4}]}{\sqrt{2m}\,\Gamma[\frac{1}{4}]}\right)^{\frac{4}{3}} \left(n+\frac{1}{2}+\frac{R(n,a)}{\pi}\right)^{\frac{4}{3}} \tag{45}$$

and gives the exact quantum values.

### *3. Conclusions.*

In conclusion, we have shown that by means of the QHJ approach it is possible to modify the WKB expressions, when incorrect, obtaining exact WKB-like formulae for the energy levels of quantum systems. This completes and extends the results found in the quoted references [4-7], where it was shown that by means of the QHJE, it is possible to give exact WKB-like expressions for the wave functions.